\documentclass[conference]{IEEEtran}
\usepackage[T1]{fontenc}
\usepackage[utf8]{inputenc}
\usepackage{amssymb}
\usepackage[pdftex]{graphicx}
\usepackage{subcaption} 
\usepackage{amsmath}
\usepackage{mathtools}
\usepackage{array}
\usepackage[noadjust]{cite}
\usepackage{algorithm}
\usepackage{algpseudocode}
\usepackage{xcolor}
\usepackage{tikz}
\usepackage{pgfplots}
\usepackage[hidelinks]{hyperref}
\usepackage[acronym]{glossaries}
\usepackage{dutchcal}
\pgfdeclarelayer{background}
\pgfdeclarelayer{foreground}
\pgfsetlayers{background,main,foreground}
\usepgfplotslibrary{fillbetween}
\usepgfplotslibrary{external}
\usetikzlibrary{spy,backgrounds}

\makeatletter
\let\MYcaption\@makecaption
\makeatother

\usepackage{subcaption}

\makeatletter
\let\@makecaption\MYcaption
\makeatother
\algnewcommand{\LineComment}[1]{\(\triangleright\) #1}

\newacronym{SNR}{SNR}{signal-to-noise ratio}
\newacronym{AWGN}{AWGN}{additive white Gaussian noise}
\newacronym{LDPC}{LDPC}{low-density parity-check}
\newacronym{QAM}{QAM}{quadrature amplitude modulation}
\newacronym{PSK}{PSK}{phase-shift keying}
\newacronym{ASK}{ASK}{amplitude-shift keying}
\newacronym{BICM}{BICM}{bit-interleaved coded modulation}
\newacronym{PAS}{PAS}{probabilistic amplitude shaping}
\newacronym{DM}{DM}{distribution matching}
\newacronym{BCE}{BCE}{binary cross-entropy}
\newacronym{LLR}{LLR}{log-likelihood ratio}
\newacronym{CQAM}{CQAM}{circular QAM}
\newacronym{BMI}{BMI}{bit-wise mutual information}
\newacronym{KL}{KL}{Kullback–Leibler}
\newacronym{BMD}{BMD}{bit-metric decoding}
\newacronym{SE}{SE}{spectral efficiency}
\newacronym{BER}{BER}{bit error rate}
\newacronym{NN}{NN}{neural network}
\newacronym{GS}{GS}{geometric shaping}
\newacronym{RBF}{RBF}{Rayleigh block fading}
\newacronym{MB}{MB}{Maxwell-Boltzmann}
\newacronym{iid}{i.i.d.\@}{independent and identically distributed}
\newacronym{SGD}{SGD}{stochastic gradient descent}
\newacronym{wrt}{w.r.t.\@}{with respect to}
\newacronym{MAP}{MAP}{maximum a posteriori}
\newacronym{LMMSE}{LMMSE}{linear minimum mean square error}
\renewcommand{\vec}[1]{\mathbf{#1}}
\newcommand{\vecs}[1]{\boldsymbol{#1}}

\newcommand{\cv}{\vec{c}}

\newcommand{\lv}{\vec{l}}

\newcommand{\xv}{\vec{x}}
\newcommand{\yv}{\vec{y}}

\newcommand{\thetav}{\vecs{\theta}}

\newcommand{\Cc}{{\cal C}}

\newcommand{\Lc}{{\cal L}}

\newcommand{\CC}{\mathbb{C}}

\newcommand{\RR}{\mathbb{R}}

\newcommand{\LB}{\left(}
\newcommand{\RB}{\right)}

\newcommand{\LSB}{\left[}
\newcommand{\RSB}{\right]}

\renewcommand{\log}[1]{\mathop{\mathrm{log}}\LB #1\RB}
\renewcommand{\exp}[1]{\mathop{\mathrm{exp}}\LB #1\RB}

\newcommand{\EE}{{\mathbb{E}}}
\newcommand{\Expect}[2]{\EE_{#1}\LSB #2\RSB}

\begin{document}
\title{Joint Learning of Probabilistic and Geometric Shaping for Coded Modulation Systems}

\IEEEoverridecommandlockouts 

\author{\IEEEauthorblockN{Fay\c{c}al Ait Aoudia and Jakob Hoydis}
\IEEEauthorblockA{Nokia Bell Labs, Paris, France\\
 \{faycal.ait\_aoudia, jakob.hoydis\}@nokia-bell-labs.com}}

\maketitle

\begin{abstract}
	We introduce a trainable coded modulation scheme that enables joint optimization of the \gls{BMI} through probabilistic shaping, geometric shaping, bit labeling, and demapping for a specific channel model and for a wide range of \glspl{SNR}.
	Compared to \gls{PAS}, the proposed approach is not restricted to symmetric probability distributions, can be optimized for any channel model, and works with any code rate $k/m$, $m$ being the number of bits per channel use and $k$ an integer within the range from $1$ to $m-1$.
	The proposed scheme enables learning of a continuum of constellation geometries and probability distributions determined by the \gls{SNR}.
	Additionally, the \gls{PAS} architecture with \gls{MB} as shaping distribution was extended with a \gls{NN} that controls the \gls{MB} shaping of a \gls{QAM} constellation according to the \gls{SNR}, enabling learning of a continuum of \gls{MB} distributions for \gls{QAM}.
	Simulations were performed to benchmark the performance of the proposed joint probabilistic and geometric shaping scheme on \gls{AWGN} and mismatched \gls{RBF} channels.
	

\end{abstract}
\glsresetall

\section{Introduction}

Constellation shaping refers to the optimization of the transmitted signal distribution to maximize the information rate.
This directly follows from the definition of the channel capacity as $C \coloneqq \max_{p(x)} I(X;Y)$, where $I(X;Y)$ is the mutual information of the channel input $X$ and output $Y$, and $p(x)$ is the distribution over the channel input.
Typical communication systems involve well-known constellation geometries such as \gls{QAM}, \gls{ASK}, and \gls{PSK}, with uniform probabilities of occurrence of the individual constellation points.
Shaping of constellations can be achieved by improving either the locations of the points or their probabilities of occurrence, referred to as geometric and probabilistic shaping, respectively.

Probabilistic shaping was shown to enable higher communication rates as well as smooth adaptation of the \gls{SE}~\cite{7307154} compared to the coarse granularity imposed by traditional schemes. These latter schemes can only operate at a fixed number of \glspl{SE} determined by the available code rates and modulation orders.
However, probabilistic shaping requires \gls{DM} to reversibly map an incoming stream of independent and uniformly distributed bits to a stream of \emph{matched} bits, such that, once modulated to channel symbols, the probabilities of occurrence of these symbols match a target distribution.
The integration of probabilistic shaping into practical \gls{BICM} systems~\cite{669123} is not straightforward.
On the one hand, performing \gls{DM} before channel coding leads to suboptimal rates, as coding alters the bit distribution.
On the other hand, performing channel coding and then \gls{DM} results in high error rates as matching breaks the code structure.
\Gls{PAS}~\cite{7307154} was introduced to integrate probabilistic shaping with \gls{BICM} systems at reasonable complexity.
However, \gls{PAS} assumes that the target distribution is symmetric around the origin.
While this holds for the \gls{AWGN} channel, it might not be necessarily the case for other channel models.
Moreover, \gls{PAS} is only compatible with code rates of $\frac{m-2}{m}$ (in 2D) or higher, $m$ being the number of bits per channel use.
Some approaches that partly alleviate this constraint were recently proposed~\cite{gultekin2019partial}.
However, they come at the cost of higher implementation complexity.

\begin{figure*}
	\centering
	\includegraphics[scale=0.38]{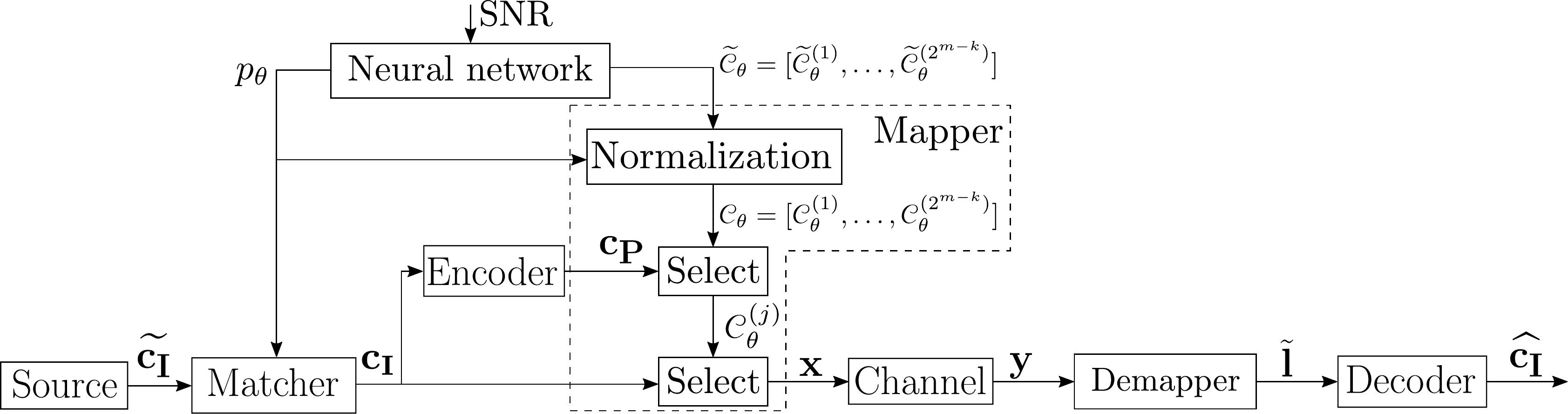}
	\caption{End-to-end system architecture\label{fig:e2e}}
\end{figure*}

In this work, a new architecture for \gls{BICM} systems is proposed which leverages the recent idea of trainable communication systems~\cite{8054694}.
The proposed approach makes no assumption on the target distribution and is compatible with any code rate $\frac{k}{m}$, $k$ being an integer within the range from $1$ to $m-1$.
Furthermore, by optimizing the proposed end-to-end system made of the modulator, channel, and demapper on the \gls{BMI}~\cite{cammerer2019trainable}, which is an achievable rate for \gls{BICM} systems~\cite{bocherer14}, joint optimization of geometric shaping, probabilistic shaping, bit labeling, and demapping is performed for a given channel model, code rate, and for a wide range of \glspl{SNR}.
Moreover, the conventional \gls{PAS} architecture with \gls{QAM} and \gls{MB} as shaping distribution was extended with a \gls{NN} that controls the \gls{MB} shaping according to the \gls{SNR}.
This enables joint optimization of the \gls{MB} distribution and demapper for any channel model and for a wide range of \glspl{SNR}.
Being able to optimize the shaping distribution $p(x)$ for any channel model is an important benefit as finding the optimal shaping distribution using conventional approaches is typically a difficult problem even when assuming knowledge of the channel distribution $p(y|x)$.

Simulations were performed on the \gls{AWGN} and mismatched \gls{RBF} channels.
On the \gls{AWGN} channel and when considering a code rate compatible with \gls{PAS}, results show that the proposed scheme achieves performance competitive with \gls{QAM} shaped with the \gls{MB} distribution using the extended \gls{PAS} architecture.
For a lower code rate that \gls{PAS} does not support, the proposed scheme achieves higher rates than 
geometric shaping and unshaped \gls{QAM}.
On the \gls{RBF} channel, joint probabilistic and geometric shaping leads to higher rates than \gls{QAM} shaped with \gls{MB} as well as geometric shaping.

The rest of this paper is organized as follows.
Section~\ref{sec:archi} details the architecture of the proposed system as well as the joint probabilistic and geometric shaping optimization algorithm.
The numerical results are presented in Section~\ref{sec:eval}.
Section~\ref{sec:conclu} concludes this paper.

\section{Joint geometric and probabilistic shaping}
\label{sec:archi}

Let us denote by $n$ the code length and by $r$ the code rate.
It is assumed that $r = \frac{k}{m}$, where $2^m$ is the modulation order, and $k \in \{1,\dots,m-1\}$.
We denote by $q = \frac{n}{m}$ the number of channel symbols per codeword, with $n$ assumed to be divisible by $m$.
The overall architecture of the proposed system is shown in Fig.~\ref{fig:e2e}.
A source generates independent and uniformly distributed bits $\boldsymbol{\widetilde{c}_I}$, which are fed to a matcher that implements a \gls{DM} algorithm~\cite{7322261}.
Given a target shaping distribution $p_{\thetav}$ of dimension $2^k$ provided by an \gls{NN} with trainable parameters denoted by $\thetav$, the matcher maps $\boldsymbol{\widetilde{c}_I}$ to a bit vector $\boldsymbol{c_I} = [\mathbf{b_I^{(1)}}~\dots~\mathbf{b_I^{(q)}}]$ of length $rn$, where $\mathbf{b_I^{(i)}},i \in \{1,\dots,q\}$, are bit vectors of dimension $k$ distributed according to $p_{\thetav}$ (note that the distribution $p_{\thetav}$ is over the bit vectors themselves and not over the individual bits).
Because of the shaping redundancy introduced by the matcher, $\boldsymbol{\widetilde{c}_I}$ has a smaller length than $rn$.
The matched information bits $\boldsymbol{c_I}$ are fed to a channel encoder that generates a vector of parity bits $\boldsymbol{c_P} = [\mathbf{b_P^{(1)}}~\dots~\mathbf{b_P^{(q)}}]$ of length $(1-r)n$, where $\mathbf{b_P^{(i)}},i \in \{1,\dots,q\}$, are bit vectors of dimension $m-k$.
Assuming a systematic code, the codeword $\cv = [\boldsymbol{c_I}~\boldsymbol{c_P}]$ is mapped to a vector of channel symbols $\xv \in \CC^q$ by a mapper and according to a constellation $\widetilde{\Cc}_{\thetav}$ also provided by the \gls{NN} with trainable parameters $\thetav$.
The constellation $\widetilde{\Cc}_{\thetav}$ consists of a set of $2^m$ points in the complex plane, numbered from $0$ to $2^m-1$.
The mapping operation will be explained in detail in Section~\ref{sec:mapping}.

On the receiver side, a differentiable demapper computes \glspl{LLR} $\tilde{\lv} \in \RR^n$ from the received symbols $\yv \in \CC^q$.
The \glspl{LLR} are fed to a decoding algorithm that reconstructs the matched information bits.
Because \gls{DM} is reversible, the unmatched information bits can be retrieved.

\subsection{Mapper architecture}
\label{sec:mapping}

The mapper maps each codeword $\cv = [\boldsymbol{c_I}~\boldsymbol{c_P}]$ to a vector of channel symbols $\xv \in \CC^q$ by mapping each bit vector $[\mathbf{b_I^{(i)}}~\mathbf{b_P^{(i)}}], i \in \{1,\dots,q\}$, of length $m$ to a channel symbol $x \in \CC$ according to the constellation $\widetilde{\Cc}_{\thetav}$.
The key idea behind the mapper architecture is to partition the constellation $\widetilde{\Cc}_{\thetav}$ into $2^{m-k}$ sub-constellations each of size $2^k$, i.e., $\widetilde{\Cc}_{\thetav} = [\widetilde{\Cc}_{\thetav}^{(1)},\dots,\widetilde{\Cc}_{\thetav}^{(2^{m-k})}]$.
All sub-constellations are normalized
\begin{equation}
	\Cc_{\thetav}^{(i)} = \frac{\widetilde{\Cc}_{\thetav}^{(i)}}{\sqrt{\sum_{x \in \widetilde{\Cc}_{\thetav}^{(i)}}p_{\thetav}(x)|x|^2}} \label{eq:norm}
\end{equation}
to form the normalized constellation $\Cc_{\thetav} = [\Cc_{\thetav}^{(1)},\dots,\Cc_{\thetav}^{(2^{m-k})}]$.
In order to map a vector of bits $[\mathbf{b_I^{(i)}}~\mathbf{b_P^{(i)}}], i \in \{1,\dots,q\}$ to a channel symbol, the sub-constellation $\Cc_{\thetav}^{(j)}$ such that $j$ has $\mathbf{b_P^{(i)}}$ as binary representation is chosen.
Next, the channel symbol $x_k \in \Cc_{\thetav}^{(j)}$ such that $k$ has $\mathbf{b_P^{(i)}}$ as binary representation is selected to modulate the bit vector $[\mathbf{b_I^{(i)}}~\mathbf{b_P^{(i)}}]$, as illustrated in Fig.~\ref{fig:e2e}.
Note that the bit vectors $\mathbf{b_I^{(i)}}$ are shaped according to $p_{\thetav}$, whereas the parity bits $\mathbf{b_P^{(i)}}$ are not as channel encoding does not preserve shaping.
The parity bits are assumed to be uniform and \gls{iid}~\cite{7307154}.
Therefore, using the proposed mapper, the constellation consists of a set of $2^{m-k}$ sub-constellations, all probabilistically shaped according to $p_{\thetav}$, but with possibly different geometries.
The normalization~(\ref{eq:norm}) ensures the power constraint $\EE\left[|x|^2\right] = 1$.

This approach is similar to \gls{PAS}~\cite{7307154}, in which $k = m-2$ and the signs of the constellation point components are selected according to the two parity bits, leading to symmetric constellations.
Our approach can be seen as a generalization of \gls{PAS}, where $k$ can take any value from the range from $1$ to $m-1$, and the sub-constellations can be freely optimized, subject only to a power constraint.

\subsection{System training}
\label{sec:training}

\begin{figure}
	\centering
	\includegraphics[scale=0.38]{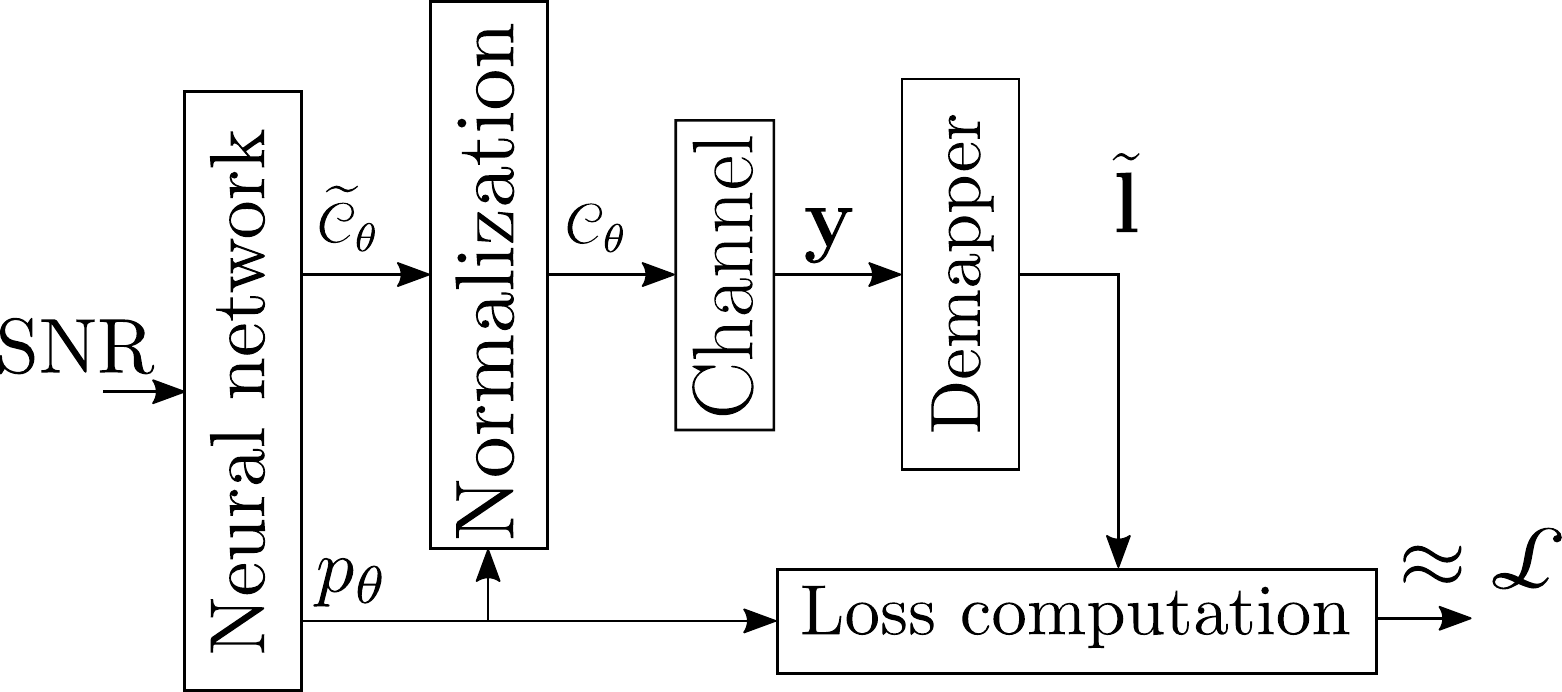}
	\caption{System architecture at training\label{fig:e2e_tr}}
\end{figure}

At training, only the mapper, channel, and demapper are considered, as channel encoding and decoding are not required to perform probabilistic and geometric shaping on the \gls{BMI}.
The system considered for training is shown in Fig.~\ref{fig:e2e_tr}.
Joint probabilistic and geometric shaping consists of jointly optimizing $p_{\thetav}$ and $\Cc_{\thetav}$ to maximize the \gls{BMI}, i.e., to find the vector of parameters that solves
\begin{align}
	\underset{\thetav}{\operatorname{argmax}}~~&R(\thetav) \label{eq:obj}\\
	\text{subject to}~~&\sum_{x\in\Cc(\thetav)}p_{\thetav}(x)|x|^2 = 1 \label{eq:cst}
\end{align}
where (\ref{eq:cst}) is an average power constraint, ensured by the normalization step performed by the mapper.
Optimizing on the \gls{BMI} is relevant as it is an achievable rate for \gls{BICM} systems~\cite{bocherer14}.
The \gls{BMI} is defined as
\begin{equation} \label{eq:bmi}
	R(\thetav) \coloneqq \left[ H_{\thetav}(X) - \sum_{i=1}^m H_{\thetav}(B_i|Y) \right]^+
\end{equation}
where $[\cdot]^+ \coloneqq \max{(\cdot,0)}$, $X$ is the random variable corresponding to the channel input, $Y$ the random variable corresponding to the channel output, and $B_1,\dots,B_m$ the random variables corresponding to the bits transmitted in a single channel use (which consist of information and parity bits).
The first term in (\ref{eq:bmi}) is the entropy of the channel input $X$,
\begin{equation}
	\label{eq:entropy}
	H_{\thetav}(X) = -\sum_{x \in \Cc_{\thetav}^{(1)}} p_{\thetav}(x) \log{p_{\thetav}(x)} + (m-k)
\end{equation}
and depends only on the shaping distribution $p_{\thetav}$.
In~(\ref{eq:entropy}), any sub-constellation $\Cc_{\thetav}^{(i)},~i \in \{1,\dots,2^{m-k}\}$, could be used in the sum as they all share the same probabilistic shaping $p_{\thetav}$.
The second term in (\ref{eq:bmi}) is the sum of the transmitted bit entropies conditioned on the channel output $Y$, i.e.,
\begin{equation} \label{eq:cch}
	H_{\thetav}(B_i|Y) = -\EE_{b_i,y} \left[\log{p(b_i|y)} \right]
\end{equation}
where $p(b_i|y)$ is the posterior probability of $b_i$ given $y$.
Note that this conditional entropy depends on $\thetav$ through the distribution of $Y$ and $B_i$.

Finding a local solution to~(\ref{eq:obj})-(\ref{eq:cst}) is done by performing \gls{SGD} on a loss function that serves as a proxy for $R$. 
As noticed in~\cite{cammerer2019trainable}, the \gls{BMI} is closely related to the total \gls{BCE} defined by
\begin{equation}
	\widehat{\Lc}(\thetav) \coloneqq -\sum_{j=1}^m \Expect{y,b_i}{\log{\widetilde{p}(b_i|y)}}
\end{equation}
where $\widetilde{p}(b_i|y)$ is the posterior probability of $b_i$ given $y$ \emph{approximated by the demapper}.
One can rewrite $\widehat{\Lc}$ as
\begin{align}
	\widehat{\Lc}(\thetav) &= \sum_{i=1}^m H_{\thetav}(b_i|y) + \sum_{i=1}^m \Expect{y}{\text{D}_{\text{KL}} \LB p(b_i|y)||\widetilde{p}(b_i|y) \RB }\\
	&= H_{\thetav}(X) - R(\thetav) + \sum_{i=1}^m \Expect{y}{\text{D}_{\text{KL}} \LB p(b_i|y)||\widetilde{p}(b_i|y) \RB }
\end{align}
where $\text{D}_{\text{KL}}$ is the \gls{KL} divergence.
Because we are performing probabilistic shaping, training on $\widehat{\Lc}$, as done in, e.g.,~\cite{cammerer2019trainable}, would lead to the minimization of $H_{\thetav}(X)$, which is not desired.
To avoid this issue, we train on the loss
\begin{align}
	\Lc(\thetav) &\coloneqq \widehat{\Lc}(\thetav) - H_{\thetav}(X)\\
				&= -R(\thetav) + \sum_{i=1}^m \Expect{y}{\text{D}_{\text{KL}} \LB p_i \LB b_i|y \RB || \widetilde{p} \LB b_i|y \RB \RB}. 	\label{eq:loss}
\end{align}
Therefore, by minimizing $\Lc$, one maximizes the \gls{BMI}.
Moreover, assuming the demapper is trainable, it would be jointly optimized with the transmitter to minimize its \gls{KL} divergence to the true posterior distribution of $B_i$ given $Y$.

\begin{figure*}
	\begin{subfigure}[b]{0.30\textwidth}
		\centering
		\includegraphics[scale=0.85]{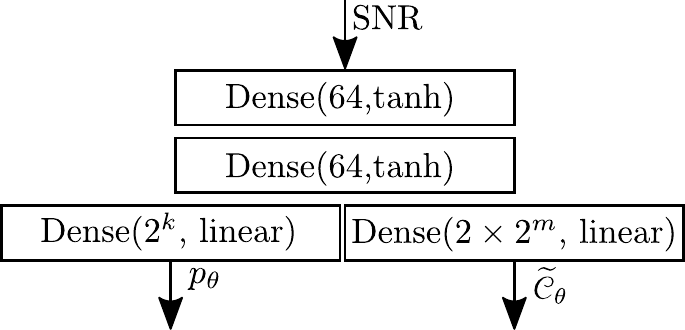}
		\caption{\gls{NN} for the PS-GS scheme\label{fig:archi_psgs}}
	\end{subfigure}
	\begin{subfigure}[b]{0.25\textwidth}
		\centering
		\includegraphics[scale=0.85]{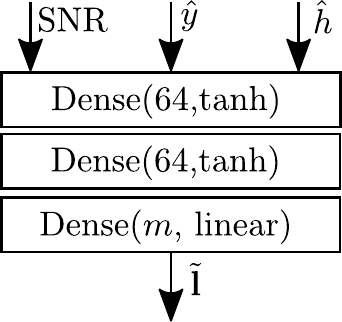}
		\caption{Demapper for the mismatched \gls{RBF} channel\label{fig:archi_demap}}
	\end{subfigure}
	\begin{subfigure}[b]{0.40\textwidth}
		\centering
		\includegraphics[scale=0.40]{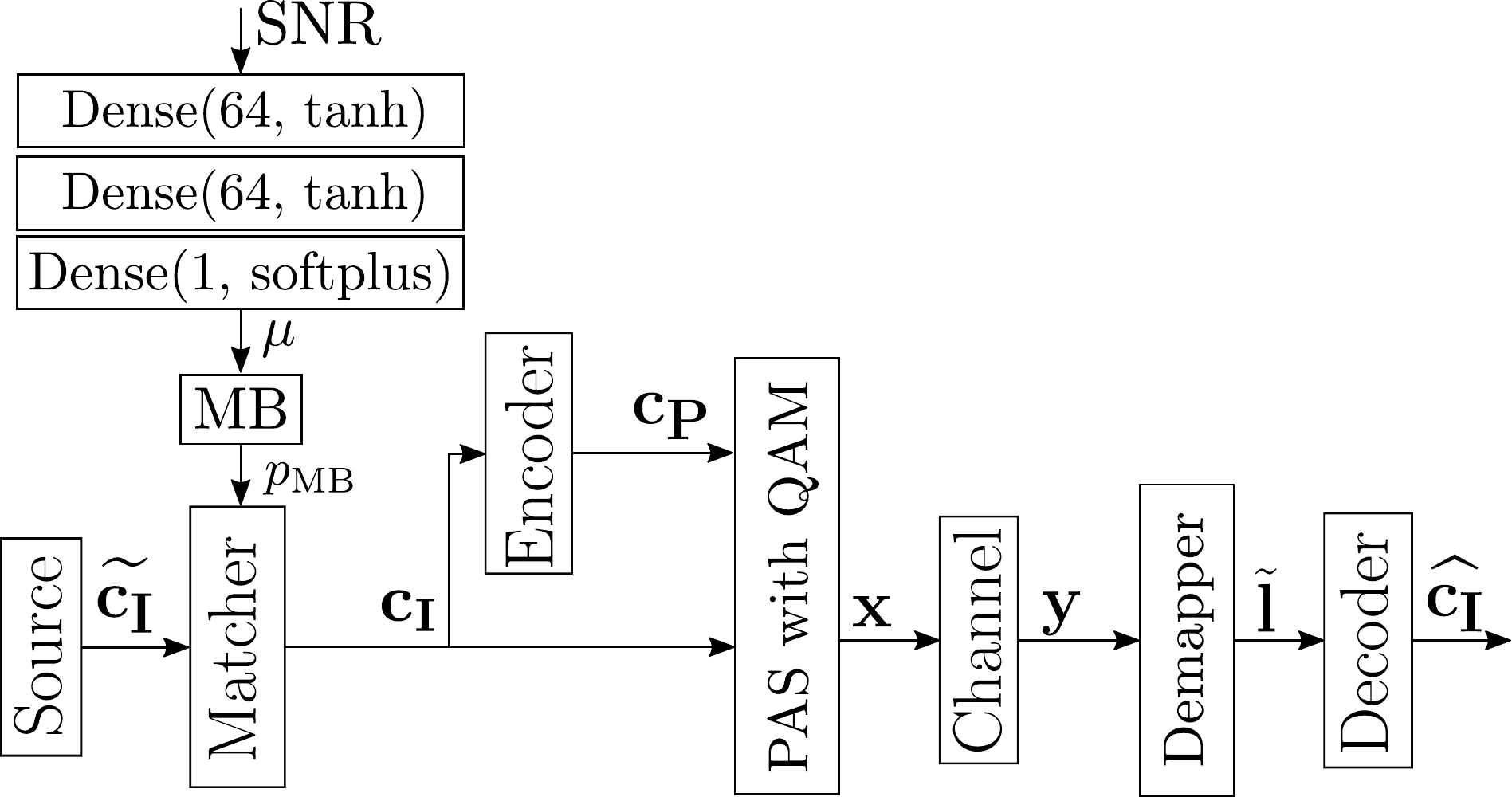}
		\caption{Extension of the \gls{PAS} architecture\label{fig:archi_deeppas}}
	\end{subfigure}
	\caption{Architectures of the \glspl{NN} used for evaluation}
\end{figure*}

A challenge of optimizing the shaping distribution with \gls{SGD} is to compute the gradient of $\Lc$ \gls{wrt} to $p_{\thetav}$.
This difficulty was addressed in~\cite{stark2019joint} by leveraging the Gumbel-Softmax trick~\cite{jang2016categorical} to implement a trainable sampling mechanism as the source of information bits.
However, the Gumbel-Softmax trick requires extra hyper-parameters to be set, and can lead to numerical instabilities at training.
In this work, we avoid the need of implementing a trainable sampling mechanism by implementing the loss in a different manner at training.
More precisely, let us denote by $\mathbcal{h}$ the channel vector state, which captures all the random elements of the channel, e.g., the noise realization of an \gls{AWGN} channel or the fading coefficient and noise realization of a fading channel.
Then, without loss of generality, (\ref{eq:cch}) can be rewritten as
\begin{multline}
	H(B_i|Y) = -\EE_{b_i,x,\mathbcal{h}} \left[\log{p\left(b_i|y(x,\mathbcal{h})\right)} \right]\\
			 = -\EE_{\mathbcal{h}} \Bigg[ \sum_{j=1}^{2^{m-k}}\sum_{x \in \Cc_{\thetav}^{(j)}} \sum_{b_i \in \{0,1\}} p(b_i|x)p_{\thetav}(x)\cdot\\
			 \log{p\left(b_i|y(x,\mathbcal{h})\right)} \Bigg] \label{eq:ch_ext}
\end{multline}
where the last equality comes from the fact that $\mathbcal{h}$ is independent of $x$ and $b_i$.
In (\ref{eq:ch_ext}), $p(b_i|x)$ is set by the labeling of the constellation point, and equals to either 0 or 1.
More precisely, each point $x$ of the constellation $\Cc_{\thetav}$ is associated to a bit vector of size $m$.
In the rest of this work, the labeling of the constellation is set to natural labeling, i.e., the first element of $\Cc_{\thetav}$ is associated to 0, the second one to 1, etc.
In other words, the bit vector associated to each constellation point is fixed, however the points can freely move within the complex plane during the optimization process, being only subject to the power constraint (\ref{eq:cst}).
When optimizing the constellation geometry $\Cc_{\thetav}$, i.e., the point positions, on the \gls{BMI}, their corresponding labeling is considered as it impacts the \gls{BMI} value.
As a result, joint optimization of the constellation geometry and labeling is achieved.

The key idea to enable optimization of the shaping distribution is to sample $\mathbcal{h}$ to estimate the outer expectation in~(\ref{eq:ch_ext}), but to explicitly implement the inner expectation over $x$ and $b_i$ 
This is not as in, e.g.,~\cite{cammerer2019trainable}, where both the channel and the source are sampled.
This trick avoids the need for a trainable sampling mechanism as the gradient of $\Lc$ is correctly computed with respect to $p_{\thetav}$.
In practice, at training, each batch example consists of sampling a channel realization $\mathbcal{h}$, and transmitting all the points forming the constellation $\Cc_{\thetav}$, as shown in Fig.~\ref{fig:e2e_tr}.
The loss $\Lc$ is then estimated by
\begin{multline}
	\Lc(\thetav) \approx -\Bigg( H_{\thetav}(X) + \sum_{i=1}^m \frac{1}{B} \sum_{l=1}^B \sum_{j=1}^{2^{m-k}} \sum_{x \in \Cc_{\thetav}^{(j)}} p_{\thetav}(x) \cdot\\
	\log{\widetilde{p}\left(b_i|y\left(x,\mathbcal{h}^{(l)}\right)\right)} \Bigg) \label{eq:l_est}
\end{multline}
where $B$ is the batch-size, and $\mathbcal{h}^{(l)}$ the $l^{\text{th}}$ sample of channel realization.

\section{Numerical results}
\label{sec:eval}

To assess the performance of the proposed scheme, referred to as PS-GS as it performs joint probabilistic and geometric shaping, we compared it to the conventional unshaped \gls{QAM}, to an optimized \gls{GS} baseline with no probabilistic shaping, and to \gls{QAM} shaped according to a \gls{MB} distribution by \gls{PAS}, referred to as MB-QAM.
For fairness, the \gls{PAS} architecture~\cite{7307154} was extended with an \gls{NN} that computes the \gls{MB} shaping from the \gls{SNR}, in order to find an optimized \gls{MB} distribution for each \gls{SNR} value (see Section~\ref{sec:pas_ext}).
The \gls{GS} baseline consists of an \gls{NN}-based mapper with uniform probability of occurrence of the points, as in, e.g.,~\cite{cammerer2019trainable}.

To evaluate the \gls{BMI} achieved by the different schemes, no \gls{DM} algorithm was used, and the information bit vectors were sampled according to the shaping distributions (or to the uniform distribution in the cases of uniform \gls{GS} and uniform \gls{QAM}).
This is equivalent to assuming the use of a perfect \gls{DM} algorithm.
\gls{AWGN} and mismatched \gls{RBF} channels were considered (see Section~\ref{sec:awgn} and~\ref{sec:rbf}, respectively).
The number of bits per channel use was set to $m = 6$ and the \gls{SNR} was defined as $1/N_0$, where $N_0$ is the noise spectral density.
The \gls{SE} was defined as the number of information bits transmitted per channel use, formally $\text{SE} \coloneqq H(X) - m(1-r)$.
Regarding the proposed PS-GS approach, two code rates were considered, $1/2$ and $2/3$.
The code rate constrains the partitioning of the constellation $\widetilde{\Cc}_{\thetav}$ in the mapper as well as the size of the shaping distribution $p_{\thetav}$ (see Section~\ref{sec:archi}).
To see how gains in \gls{BMI} translate into gains in \gls{BER}, a standard IEEE 802.11n code was used, with a length of $n = 1944$ bits and a code rate of $2/3$.
This code rate was chosen as it is compatible with \gls{PAS} for $m=6$.
However, \gls{PAS} cannot operate with a code rate of $1/2$.
A conventional sum-product belief-propagation decoder with 100 iterations was leveraged for channel decoding.

Regarding the proposed scheme, the \gls{NN} that generates the shaping distribution $p_{\thetav}$ and the constellation $\widetilde{\Cc}_{\thetav}$ was made of two dense layers with 64 units each and hyperbolic tangent activation function, followed by two parallel dense layers with linear activation, one with $2^k$ units to generate the shaping distribution, and the other with $2^{m+1}$ units to generate the real and imaginary parts of the constellation points.
This architecture is shown in Fig.~\ref{fig:archi_psgs}.
When considering the \gls{AWGN} channel, the true posterior distribution on bits was implemented by the demapper.
For the mismatched \gls{RBF} channel, as no efficient implementation of the optimal demapper is known to the best of our knowledge, the demapper was implemented by an \gls{NN} whose architecture is shown in Fig.~\ref{fig:archi_demap}.
The \gls{NN}-based demapper was jointly optimized with the transmitter.
In Fig.~\ref{fig:archi_demap}, $\hat{y}$ is the equalized received symbol and $\hat{h}$ the \gls{LMMSE} channel estimate.
For fairness, an \gls{NN}-based demapper was leveraged for all considered schemes when considering the mismatched \gls{RBF} channel, including the \gls{QAM} baseline for which the transmitter includes no trainable parameters.

Training was done with batches of size $1000$, using the Adam optimizer~\cite{Kingma15} with the learning rate set to $10^{-3}$, and by uniformly sampling the \gls{SNR} from the range from 0 to 20\,dB for the \gls{AWGN} channel, and from the range from 5 to 25\,dB for the mismatched \gls{RBF} channel.
For all the schemes that required training, the best of five seeds is shown.

\subsection{Extending PAS with deep learning}
\label{sec:pas_ext}

For fairness, the \gls{PAS} architecture with \gls{MB} as shaping distribution and \gls{QAM} geometry was extended with an \gls{NN} that controls the \gls{MB} shaping according to the \gls{SNR}.
More precisely, if we denote by $\Cc_{\text{QAM}}$ the \gls{QAM} constellation vector, the probability of occurrence of a point $x \in \Cc_{\text{QAM}}$ using \gls{MB} shaping is
\begin{equation}
	p_{\text{MB}}(x) = \frac{\exp{-\mu |x|^2}}{\sum_{x'\in\Cc_{\text{QAM}}} \exp{-\mu |x'|^2}}
\end{equation}
where $\mu \geq 0$ controls the constellation shaping.
The \gls{PAS} architecture was extended with an \gls{NN} that computes $\mu$ from the \gls{SNR}, as shown in Fig.~\ref{fig:archi_deeppas}.
This architecture is similar to the one used for joint probabilistic and geometric shaping (see Fig.~\ref{fig:e2e}), except that the mapper is implemented following the \gls{PAS} scheme.
Extending \gls{PAS} as proposed in this work enables the learning of a continuum of \gls{MB} distributions determined by the \gls{SNR} for any channel model, constellation geometry, and differentiable demapper.
Training is done using the approach described in Section~\ref{sec:training}, i.e., by transmitting for each channel realization the entire \gls{QAM} constellation, and by performing \gls{SGD} on the loss~(\ref{eq:l_est}).

\subsection{AWGN channel}
\label{sec:awgn}

\begin{figure}

	\begin{subfigure}{\linewidth}
		\begin{tikzpicture}
			\begin{axis}[
				grid=both,
				grid style={line width=.4pt, draw=gray!10},
				major grid style={line width=.2pt,draw=gray!50},
				xlabel={$E_s/N_0$ [dB]},
				ylabel={BMI [bit / channel use]},
				legend columns=3,
				legend style={font=\small, at={(0.5,0.95)}, anchor=south},
                xlabel style={font=\small},
                ylabel style={font=\small},
                xlabel near ticks,
                tick label style={font=\small},
				xmin=-1,
				xmax=21,
			]
			
				\addplot[thick, dotted, black] table [x=es_n0, y=rate, col sep=comma] {figs/cap_awgn.csv};	
				\addplot[thick, black, mark=o, mark repeat=10] table [x=es_n0, y=rate, col sep=comma] {figs/psgs0.66_awgn.csv};
				\addplot[thick, blue, mark=diamond, mark repeat=10] table [x=es_n0, y=rate, col sep=comma] {figs/psqam_awgn.csv};
				\addplot[thick, red, mark=x, mark repeat=10] table [x=es_n0, y=rate, col sep=comma] {figs/psgs0.50_awgn.csv};
				\addplot[thick, olive, mark=square, mark repeat=10] table [x=es_n0, y=rate, col sep=comma] {figs/gs_awgn.csv};
				\addplot[thick, magenta, mark=triangle, mark repeat=10] table [x=es_n0, y=rate, col sep=comma] {figs/qam_awgn.csv};

				\addlegendentry{Capacity}
				\addlegendentry{PS-GS(2/3)}
				\addlegendentry{MB-QAM(2/3)}
				\addlegendentry{PS-GS(1/2)}
				\addlegendentry{GS}
				\addlegendentry{QAM}
				
			\coordinate (pt) at (axis cs:11,3.7);
			
				\end{axis}

				\node[pin=-5:{%
				    \begin{tikzpicture}[baseline,trim axis left,trim axis right]
				    \begin{axis}[
				        tiny,
					grid=both,
					grid style={line width=.4pt, draw=gray!10},
					major grid style={line width=.2pt,draw=gray!50},
				      xmin=10.8,xmax=11.2,
				      ymin=3.60,ymax=3.80,
				      enlargelimits,
				    ]

				\addplot[thick, dotted, black] table [x=es_n0, y=rate, col sep=comma] {figs/cap_awgn.csv};	
				\addplot[thick, black, mark=o, mark repeat=10] table [x=es_n0, y=rate, col sep=comma] {figs/psgs0.66_awgn.csv};
				\addplot[thick, blue, mark=diamond, mark repeat=10] table [x=es_n0, y=rate, col sep=comma] {figs/psqam_awgn.csv};
				\addplot[thick, red, mark=x, mark repeat=10] table [x=es_n0, y=rate, col sep=comma] {figs/psgs0.50_awgn.csv};
				\addplot[thick, olive, mark=square, mark repeat=10] table [x=es_n0, y=rate, col sep=comma] {figs/gs_awgn.csv};
				\addplot[thick, magenta, mark=triangle, mark repeat=10] table [x=es_n0, y=rate, col sep=comma] {figs/qam_awgn.csv};

				    \end{axis}
				    \end{tikzpicture}%
				}] at (pt) {};

		\end{tikzpicture}
		\caption{BMI. PS-GS and MB-QAM achieve similar rates.}
		\label{fig:awgn_rate}
	\end{subfigure}
	
	\begin{subfigure}{\linewidth}
		\begin{tikzpicture}
			\begin{axis}[
				grid=both,
				grid style={line width=.4pt, draw=gray!10},
				major grid style={line width=.2pt,draw=gray!50},
				xlabel={$E_s/N_0$ [dB]},
				ylabel={SE [bit / channel use]},
				legend columns=3,
				legend style={font=\small, at={(0.5,0.95)}, anchor=south},
                xlabel style={font=\small},
                ylabel style={font=\small},
                xlabel near ticks,
                tick label style={font=\small},
				xmin=-1,
				xmax=21,
			]
				\addplot[thick, black, mark=o, mark repeat=10] table [x=es_n0, y=se, col sep=comma] {figs/psgs0.66_awgn.csv};
				\addplot[thick, blue, mark=diamond, mark repeat=10] table [x=es_n0, y=se, col sep=comma] {figs/psqam_awgn.csv};
				\addplot[thick, red, mark=x, mark repeat=10] table [x=es_n0, y=se, col sep=comma] {figs/psgs0.50_awgn.csv};
  				\addplot[mark=none,very thick, domain=0:20, black, dashed, samples=2] {3};
   				\addplot[mark=none,very thick, domain=0:20, black, dashdotted, samples=2] {4};
   				
   				\node[above,right] at (axis cs: 0.0,3.1) {\small No PS(1/2)};
   				\node[above,right] at (axis cs: 0.0,4.1) {\small No PS(2/3)};
				
			\end{axis}

		\end{tikzpicture}
		\caption{SE. PS-GS and MB-QAM adapt the SE according to the \gls{SNR}.}
		\label{fig:awgn_se}
	\end{subfigure}
	
	\begin{subfigure}{\linewidth}
		\begin{tikzpicture}
			\begin{axis}[
				grid=both,
				grid style={line width=.4pt, draw=gray!10},
				major grid style={line width=.2pt,draw=gray!50},
				xlabel={$E_s/N_0$ [dB]},
				ylabel={BER},
				ymode=log,
				ymin=1e-5,
				legend columns=3,
				legend style={font=\small, at={(0.5,0.95)}, anchor=south},
                xlabel style={font=\small},
                ylabel style={font=\small},
                xlabel near ticks,
                tick label style={font=\small},
			]
				\addplot[thick, black, mark=o] table [x=es_n0, y=ber, col sep=comma] {figs/psgs0.66_awgn_ber.csv};
				\addplot[thick, blue, mark=diamond] table [x=es_n0, y=ber, col sep=comma] {figs/psqam_awgn_ber.csv};
				\addplot[thick, olive, mark=square] table [x=es_n0, y=ber, col sep=comma] {figs/gs_awgn_ber.csv};
				\addplot[thick, magenta, mark=triangle] table [x=es_n0, y=ber, col sep=comma] {figs/qam_awgn_ber.csv};


			\end{axis}

		\end{tikzpicture}
		\caption{BER achieved with a code rate of $2/3$}
		\label{fig:awgn_ber}
	\end{subfigure}

	\caption{Results for the AWGN channel\label{fig:awgn}}
\end{figure}
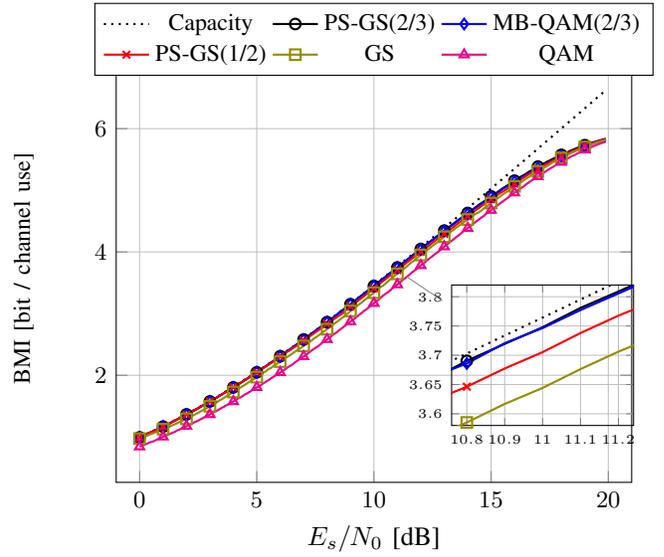
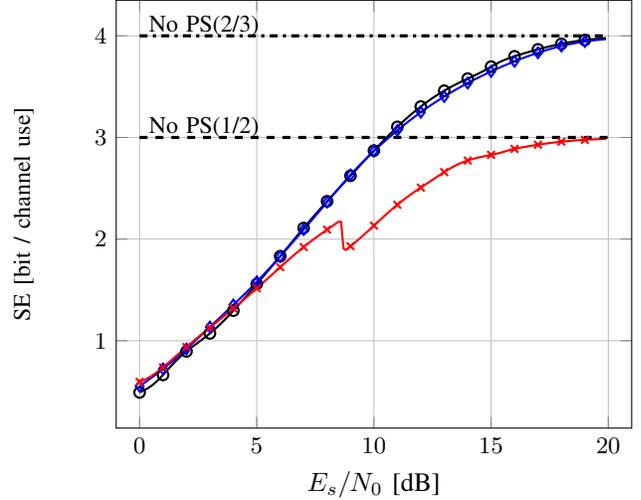
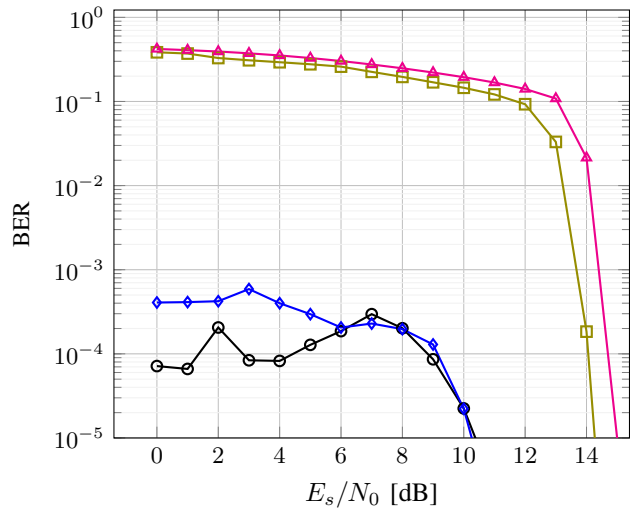

Considering the \gls{AWGN} channel, the true posterior distribution was implemented in the demapper.
Therefore, the \gls{KL} divergence in~(\ref{eq:loss}) is null, and the loss function $\Lc$ equals the \gls{BMI} $R$ up to the sign.
The \glspl{BMI} achieved by the compared schemes were estimated by Monte Carlo sampling of~(\ref{eq:loss}) and are shown in Fig.~\ref{fig:awgn_rate}.
The code rate is indicated in parenthesis in the legend.
One can see that PS-GS with a code rate of $2/3$ and MB-QAM achieve essentially the same \gls{BMI}, followed by PS-GS with a code rate of $1/2$.
The lower performance of PS-GS with a code rate of $1/2$ can be explained by the higher number of sub-constellations into which $\widetilde{\Cc}_{\thetav}$ must be partitioned by the mapper due to the lower code rate.
Indeed, the higher the number of sub-constellations into which the constellation $\widetilde{\Cc}_{\thetav}$ must be partitioned, the stronger the constraint as all sub-constellations must share the same probabilities of occurrence of points.
Note that conventional \gls{PAS} does not operate at a code rate of $1/2$ with $m$ set to 6.
However, PS-GS with a code rate of $1/2$ still achieves significantly higher rates than \gls{GS} alone, which itself outperforms uniform \gls{QAM}.

Fig.~\ref{fig:awgn_se} shows the corresponding \gls{SE}.
When considering uniform shaping, i.e., unshaped \gls{QAM} and \gls{GS} alone, the \gls{SE} is constantly equal to $4$ as the \gls{SE} only depends on the code rate and modulation order.
On the other hand, probabilistic shaping enables smooth adaptation of the \gls{SE}, as opposed to conventional systems for which only coarse adaption is possible as the \gls{SE} is determined by the available code rates and modulation orders.
From this figure, one can see that PS-GS with a code rate of $2/3$ and MB-QAM achieve similar \glspl{SE}.
PS-GS with a code rate of $1/2$ obviously achieves lower \gls{SE} due to the lower code rate.

Fig.~\ref{fig:awgn_ber} shows the \gls{BER} achieved by the compared schemes for a code rate of $2/3$.
One can see that both PS-GS and MB-QAM enable significantly lower \gls{BER} than uniform approaches.
However, this is at the cost of a significantly lower \gls{SE} (Fig.~\ref{fig:awgn_se}).
For low \glspl{SNR} (lower than $6\:$dB), PS-GS achieves lower \glspl{BER} than MB-QAM, at the cost of slightly lower \glspl{SE} at it can be seen in Fig.~\ref{fig:awgn_se}.

\subsection{Mismatched RBF channel}
\label{sec:rbf}
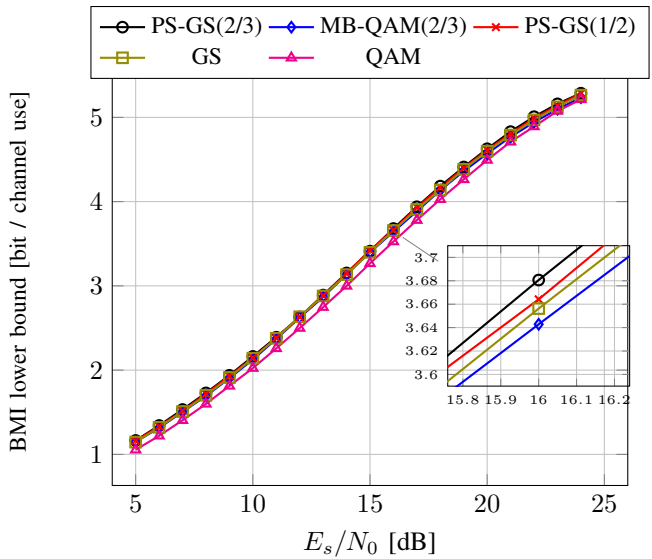
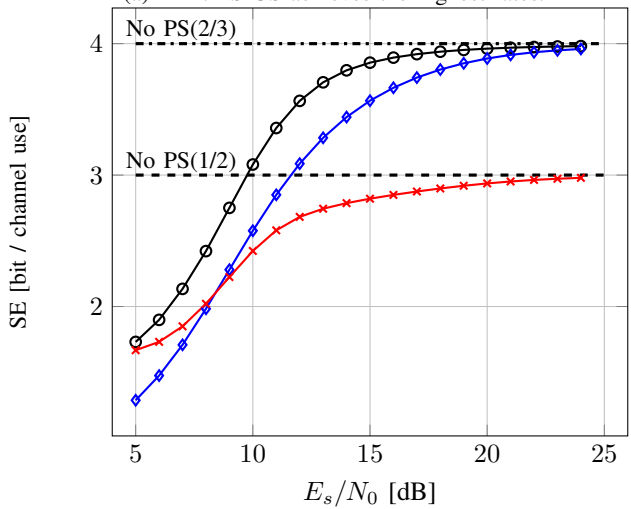
\begin{figure}
	\centering

	\begin{subfigure}{\linewidth}
		\begin{tikzpicture}
			\begin{axis}[
				grid=both,
				grid style={line width=.4pt, draw=gray!10},
				major grid style={line width=.2pt,draw=gray!50},
				xlabel={$E_s/N_0$ [dB]},
				ylabel={BMI lower bound [bit / channel use]},
				legend columns=3,
				legend style={font=\small, at={(0.5,0.95)}, anchor=south},
                xlabel style={font=\small},
                ylabel style={font=\small},
                xlabel near ticks,
				xmin=4,
				xmax=26,
			]
			
				\addplot[thick, black, mark=o] table [x=es_n0, y=rate, col sep=comma] {figs/psgs0.66_rbf.csv};
				\addplot[thick, blue, mark=diamond] table [x=es_n0, y=rate, col sep=comma] {figs/psqam_rbf.csv};
				\addplot[thick, red, mark=x] table [x=es_n0, y=rate, col sep=comma] {figs/psgs0.50_rbf.csv};
				\addplot[thick, olive, mark=square] table [x=es_n0, y=rate, col sep=comma] {figs/gs_rbf.csv};
				\addplot[thick, magenta, mark=triangle] table [x=es_n0, y=rate, col sep=comma] {figs/qam_rbf.csv};

				\addlegendentry{PS-GS(2/3)}
				\addlegendentry{MB-QAM(2/3)}
				\addlegendentry{PS-GS(1/2)}
				\addlegendentry{GS}
				\addlegendentry{QAM}
				
			\coordinate (pt) at (axis cs:16,3.65);
			
				\end{axis}

				\node[pin=-5:{%
				    \begin{tikzpicture}[baseline,trim axis left,trim axis right]
				    \begin{axis}[
				        tiny,
					grid=both,
					grid style={line width=.4pt, draw=gray!10},
					major grid style={line width=.2pt,draw=gray!50},
				      xmin=15.8,xmax=16.2,
				      ymin=3.60,ymax=3.70,
				      enlargelimits,
				    ]

				\addplot[thick, black, mark=o] table [x=es_n0, y=rate, col sep=comma] {figs/psgs0.66_rbf.csv};
				\addplot[thick, blue, mark=diamond] table [x=es_n0, y=rate, col sep=comma] {figs/psqam_rbf.csv};
				\addplot[thick, red, mark=x] table [x=es_n0, y=rate, col sep=comma] {figs/psgs0.50_rbf.csv};
				\addplot[thick, olive, mark=square] table [x=es_n0, y=rate, col sep=comma] {figs/gs_rbf.csv};
				\addplot[thick, magenta, mark=triangle] table [x=es_n0, y=rate, col sep=comma] {figs/qam_rbf.csv};

				    \end{axis}
				    \end{tikzpicture}%
				}] at (pt) {};

		\end{tikzpicture}
		\caption{BMI. PS-GS achieves the highest rates.}
		\label{fig:rbf_rate}
	\end{subfigure}
	
	\begin{subfigure}{\linewidth}
		\begin{tikzpicture}
			\begin{axis}[
				grid=both,
				grid style={line width=.4pt, draw=gray!10},
				major grid style={line width=.2pt,draw=gray!50},
				xlabel={$E_s/N_0$ [dB]},
				ylabel={SE [bit / channel use]},
				legend columns=3,
				legend style={font=\small, at={(0.5,0.95)}, anchor=south},
                xlabel style={font=\small},
                ylabel style={font=\small},
                xlabel near ticks,
				xmin=4,
				xmax=26,
			]
				\addplot[thick, black, mark=o] table [x=es_n0, y=se, col sep=comma] {figs/psgs0.66_rbf.csv};
				\addplot[thick, blue, mark=diamond] table [x=es_n0, y=se, col sep=comma] {figs/psqam_rbf.csv};
				\addplot[thick, red, mark=x] table [x=es_n0, y=se, col sep=comma] {figs/psgs0.50_rbf.csv};
  				\addplot[mark=none,very thick, domain=5:25, black, dashed, samples=2] {3};
   				\addplot[mark=none,very thick, domain=5:25, black, dashdotted, samples=2] {4};
   				
   				\node[above,right] at (axis cs: 4.2,3.1) {\small No PS(1/2)};
   				\node[above,right] at (axis cs: 4.2,4.1) {\small No PS(2/3)};
				

			\end{axis}

		\end{tikzpicture}
		\caption{SE. PS-GS and MB-QAM adapt the SE according to the \gls{SNR}.}
		\label{fig:rbf_se}
	\end{subfigure}

	\caption{Results for the mismatched RBF channel\label{fig:rbf}}
\end{figure}

Considering the mismatched \gls{RBF} channel, an \gls{NN}-based demapper was considered as no exact solution of low complexity is available.
Therefore, $\Lc$ is a lower bound on the \gls{BMI} up to the sign.
Assuming the \gls{NN} implementing the demapper is of high enough approximation capacity, it can closely approximate the true posterior distribution, making $\Lc$ a tight bound on the \gls{BMI}.

Because of space restriction, only the \gls{BMI} and \gls{SE} are shown for the mismatched \gls{RBF} channel. 
Fig.~\ref{fig:rbf_rate} shows the lower bounds on the \gls{BMI} achieved by the compared schemes and obtained by Monte Carlo sampling of~(\ref{eq:loss}).
As one can see, PS-GS with a code rate of $2/3$ achieves the highest values, followed by PS-GS with a code rate of $1/2$ and \gls{GS}.
As opposed to the \gls{AWGN} channel, these schemes outperform MB-QAM.
Looking at Fig.~\ref{fig:rbf_se}, one can see that PS-GS achieves a higher \gls{SE} than MB-QAM for a code rate of $2/3$.
The gains observed using the proposed scheme in this paper demonstrate the benefits of being able to learn the shaping for any channel model.

\section{Conclusion}
\label{sec:conclu}

We have introduced a novel \gls{NN} architecture that enables joint probabilistic and geometric shaping for \gls{BICM} systems in an end-to-end manner.
The proposed architecture is compatible with any code rate $k/m$, where $m$ is the number of bits per channel use and $k$ is an integer within the range from $1$ to $m-1$.
It further enables joint optimization of the geometric shaping, probabilistic shaping, bit labeling, and demapping on the \gls{BMI} for any channel model and for a wide range of \glspl{SNR}.
Numerical results show that the proposed approach achieves a \gls{BMI} competitive with \gls{QAM} shaped according to a \gls{MB} distribution by \gls{PAS} at a compatible code rate.
At a code rate at which \gls{PAS} does not operate, it outperforms geometric shaping alone.
On the mismatched \gls{RBF} channel, the proposed scheme achieves higher rates than shaped \gls{QAM} for all considered rates, showing the benefits of being able to learn the shaping for an arbitrary channel model.

\section*{Acknowledgment}
The authors thank Laurent Schmalen, Sebastian Cammerer, Stephan ten Brink, and Fanny Jardel for comments that have greatly improved the manuscript.

\bibliographystyle{IEEEtran}
\bibliography{IEEEabrv,bibliography}
\pagebreak
\end{document}